# Chapter 1

# Chapter 1 - Large Scale Features of Southwest Monsoon During 2015

*Hamza Varikoden, Bhupendra Bahadur Singh, K.P. Sooraj, Manish K. Joshi, B. Preethi, Milind Mujumdar and M. Rajeevan*

**1.1 Rainfall features**

During 2015, the southwest monsoon (SWM) rainfall over the country remained deficient with seasonal rainfall of about 86% of the long period average (Table 1.1). Last year, the seasonal rainfall deficiency over the country as a whole was 12% (www.imd.gov.in). Thus, this is a fourth episode of two consecutive years, with deficient monsoon, similar to 1904-05, 1965-66 and 1986-87 (www.imd.gov.in).

However, the analysis at a monthly scale reveals the intrinsic and pronounced spatial variability of the 2015 monsoon. For June 2015, rainfall over the country as a whole was 16% above the long period average (Figures 1.1a-c and Table 1.1). During this month, the rainfall activity was above normal over Central India, with a deluge of rainfall extending to eastern and southern Peninsular India. But, the monthly rainfall remained subdued over the northeastern part of India. This was in contrast to the rainfall during June 2014 (figure not shown), which received only 58% of the long period average. During July 2015 (Figures 1.1d-f), the rainfall across the country was deficient by 16% (Table 1.1). The monthly rainfall was above normal only over central to northwest India, with the rest of the country registering below normal rainfall activity during July 2015. Note that the daily cumulative rainfall over India as a whole, until 31st July (www.imd.gov.in), showed a percentage departure of only -4% from the long period cumulative mean. In contrast, the 2014 SWM registered a dauntingly enormous percentage deficit of 22% by 31 July 2014 (figure not shown). During August 2015 (Figures 1.1g-i), the country as a whole witnessed subdued rainfall activity, with the rainfall deficit deteriorating further to 24%. However, there was isolated rainfall activity over the eastern Peninsular region, particularly confined to Tamil Nadu and erstwhile Andhra Pradesh. During September 2015 (Figures 1.1j-l), the rainfall activity further reduced, with the whole country receiving only 76% of the long period average rainfall. Rainfall activity was confined to Orissa and its neighboring regions. In particular, the region extending from central to northeast India registered an acute rainfall deficit. The seasonal rainfall (Figures 1.1m-o) for the entire country was 86.4% of the long period average, thus amounting the rainfall deficit to 14%. Over the four homogenous rainfall zones (defined by IMD, www.imd.gov.in), the seasonal cumulative rainfall showed a percentage of departure -17, -8.4, -16.4 and -15.4%, respectively over the northwest, east-northeast, central and south Peninsular India. Moreover, the standardized daily rainfall anomalies over the core monsoon zone (Rajeevan et al., 2010) recorded a significant amount of break monsoon days (accumulating to 36) during the entire season (Figure 1.2), with prolonged below normal rainfall activity during August to September 2015, further corroborating the seasonal rainfall deficit across the country (Rajeevan et al., 2010). There were only isolated above normal rainfall activities (during 13-26 June and 15-23 September). The subdued rainfall activity across the country is further manifested in the weekly rainfall departure, as shown in Figure 3.3.

**1.2 Mean and anomalies of wind at lower and upper levels**

During the southwest monsoon season, the winds at lower levels from the southern hemisphere cross the equator off the Somali coast and turn towards the Indian Peninsula with an average speed of around 15 ms-1 and this wind is called the Low Level Jetstream (LLJ, Joseph and Raman, 1966). The strength of the LLJ increases from June to August and thereafter it decreases. During 2015, the LLJ was weaker compared to the climatological value (Figure 1.3). The weaker LLJ is due to a weaker Mascarene High and associated weaker outflow and cross equatorial flow along the Somali coast. The strong easterly anomalies over the Arabian Sea and BoB, indicate the weaker LLJ. Over the Indian region, anomalous anticyclonic flow is observed suggesting a weaker monsoon circulation. In July and August 2015, the Arabian Sea and south Bay of Bengal branches of the LLJ were considerably weaker. During September 2015, the low level winds over the Arabian Sea slightly intensified; however the BoB branch remained weak. For the season as a whole, the wind at 850 hPa did not pick up enough strength over the entire north Indian Ocean, showing easterly anomalies over the north Indian Ocean and weaker cross equatorial flow across the Somali coast (Figures 1.3e,j). The weaker cross equatorial flow was due to a weaker Mascarene High in the southern hemisphere and associated outflow (figure not shown). Over the equatorial Pacific, westerly anomalies were observed at 850 hPa suggesting El Niño conditions as documented in section 1.3 (Figure 1.5).

Similar to the LLJ evolution during the SWM season, the evolution of the Tropical Easterly Jet Stream (TEJ) is also a characteristic feature of the SWM (Koteswaram, 1958). It evolves over the north off-equatorial regions with slight northward movement and its core is situated at an altitude of about 150 hPa. During June 2015, the TEJ evolved around the central equatorial Indian Ocean (Figure 1.4a), however, an anomalous westerly component, over the northern Indian Ocean, indicates a weaker TEJ (Figure 1.4f). Interestingly, the Tibetan anticyclone depicted weaker to moderate monthly variation (Figures 1.4f-i), however on the seasonal scale this structure is weak (Figure 1.4j). The slight northward movement of upper level tropical easterlies (part of the TEJ) and subtropical westerlies during the month of July 2015 seems to be conducive for moderate monsoonal activity (Figure 1.4b and g). During September, TEJ was weak and located over the equator. The Subtropical Westerly Jet stream was shifted southward with westerly anomalies over the northern parts of the country. However, the large-scale anomalous circulation over Asia-Pacific region clearly indicates weaker monsoon flow. Over the equatorial Pacific Ocean, the 200 hPa wind anomalies were easterlies as seen during canonical El Niño conditions (Rasmusson and Carpenter 1983) indicating weaker/reversed Walker Circulation. In general, the weaker SWM circulation patterns are consistent with the deficient rainfall activity over India as evident from Figure 1.1.

**1.3 Sea Surface Temperature (SST) anomalies**

During the month of June, the Indo-Pacific SST anomalies were mostly positive, with excessive warming over the equatorial Indian Ocean and the eastern Pacific Ocean. The westward extension of warming beyond the dateline, over the equatorial Pacific Ocean (Figure 1.5a), is suggestive of emerging canonical El Niño conditions (Rassmussen and Carpenter, 1983). The warm anomalies are prominent over the northeast Pacific Ocean as well. The BoB and the Arabian Sea also experienced positive SST anomalies during the month of June. Meanwhile, the amplitude of the anomalous warming over the Indian Ocean reduced during the following summer monsoon months and resulted in a negative north-south anomalous SST gradient (Shankar et al., 2007), particularly over the Bay of Bengal (Figure 1.5b). This feature is somewhat surprising. Moreover, the El Niño signal became prominent during the month of July. The detailed description of El Niño related Oceanic features is given in Chapter 2. This anomalous Indo-Pacific SST pattern is consistent with the reduction in July rainfall (Figure 1.1f) and the associated weakened low level monsoon circulation (Figure 1.3g). During the month of August, the Indian Ocean witnessed a signature of moderate positive IOD (Saji et al., 1999), with anomalously warm SSTs over the western equatorial Indian Ocean and moderately cooler-than-normal SSTs over the eastern equatorial Indian Ocean (Figure 1.5c). However, the signature weakened during September due to the weakening of negative anomalies over the eastern Indian Ocean (Figure 1.5d). On the other hand, the anomalous warming over the western Indian Ocean strengthened and extended northward over the Arabian Sea. The Indian Ocean warming and Indian Ocean Dipole are dealt with details in Section 2.4 of Chapter 2.

It is interesting to note that the abnormal warming of the northeast Pacific Ocean has prevailed throughout the season (Figure 1.5a-d), strengthening and extending westward, especially from June to September. Over the tropical Pacific, the maximum SST anomalies were located along the Niño1+2 region during the beginning of the monsoon season (in June). As the season progressed, the maximum SST anomalies moved to the Niño 3 and Niño 3.4 regions (Figures 1.5b-d). The El Niño related equatorial Pacific SST gradient also strengthened with the progression of the seasons and the maximum amplitude was observed during the month of September (Figure 1.5d). Concisely, the most prominent signal for the season as a whole is the eastern Pacific El Niño pattern extending to the date line and the anomalous warming over the northeast Pacific Ocean (Figure 1.5e), which made the El Niño condition of 2015 distinct from previous El Niño years. On the other hand, the most interesting feature of the Indian Ocean is the existence of a negative north-south gradient in SST anomalies, especially in the Bay of Bengal (Shankar et al., 2007) with more anomalous warming over the western Indian Ocean as compared to that of its eastern counterpart (Figure 1.5e).

**1.4 Mean and anomalous Outgoing Longwave Radiation (OLR) patterns**

Figure 1.6 shows that during June, high convective activity with OLR less than 200 $Wm^{-2}$ (Gruber and Krueger, 1984; Lau et al., 1997) is mainly observed over the equatorial western Pacific, BoB and Arabian Sea. However, the anomaly pattern indicates enhanced convection over the Arabian Sea, but with suppressed convection over the equatorial western Pacific and BoB. This contrasting convective pattern over the Arabian Sea and the BoB is intriguing. The July mean pattern shows enhanced convection over the head BoB, but with reduced convection over the equatorial western Pacific as compared to that of June. The anomaly pattern during July shows suppressed convection over the equatorial western Pacific and most of the north Indian Ocean. A clear east-west gradient of OLR anomalies is seen in the equatorial Pacific during July. This is also consistent with the July lower level wind pattern (Figure 1.3) where a westerly anomaly is seen over the equatorial Pacific.

During August the mean convection over the BoB has weakened as compared to July and the corresponding anomaly pattern reveals suppressed convection over most of the equatorial western Pacific and north Indian Ocean. The east-west gradient in the anomalous pattern over the equatorial Pacific has strengthen (Figure 1.6h). These features are consistent with the anomalous upper level wind circulation as discussed earlier (Figure 1.4). During September, convection weakens over the BoB, while strengthening of convection over the central and eastern Pacific is evident. The anomaly pattern shows a stronger east-west gradient in the equatorial Pacific depicting a strong El Niño signal (Figure 1.5). It is also interesting to note the enhanced convection over the southern Indian Peninsula (Figure 1.6i). The extension of suppressed convection from the warm pool region to the central Indian Ocean seems to be associated with the weakening of negative SST anomalies over the eastern Indian Ocean (Figure 1.5d). The seasonal mean OLR (Figure 1.6e) shows intense convection over the BoB and the equatorial eastern Pacific, associated with the El Niño event. The seasonal anomaly clearly reflects the Pacific El Niño conditions with suppressed convection over the equatorial western Pacific, BoB and over the Indian subcontinent, while enhanced convection over the equatorial central Pacific extends up to the equatorial eastern Pacific. It is worth noting that the suppressed convection over the west Pacific warm pool region extends westwards up to the central equatorial Indian Ocean (Figure 1.6j).

**1.5 Velocity Potential anomalies at lower and upper levels**

Figure 1.7 shows the velocity potential anomalies at 850 hPa and 200 hPa for the months of June, July, August, and September as well as for the entire monsoon season. The low-level convergence (rising branch)/divergence (sinking branch) at 850 hPa is generally manifested as the upper level divergence/convergence at 200 hPa, which strongly indicates the large-scale zonal overturning circulations (Krishnamurti, 1971; Palmer et al., 1992). The climatology of velocity potential at 850 hPa (figure not shown) shows a rising branch of the Walker circulation over the western Pacific and Maritime Continent, coinciding with strong convection over the region, and a descending branch over the eastern Pacific. These ascending and descending branches are connected with low-level easterlies and upper level westerlies over the central and eastern Pacific. During June, the anomalous east-west overturning circulation over the Indo-Pacific region is not very clear. In contrast to June, the anomalies for the months of July to September as well as for the entire monsoon season depict a weakening of the zonal overturning circulation. This is consistent with the eastern Pacific warming signal (Figure 1.5), and indicates an El Niño related anomalous Walker cell, i.e., showing divergence (convergence) over the western end of Pacific extending to the Indian subcontinent and convergence (divergence) over the eastern to central Pacific at 850 hPa (200 hPa). This anomalous circulation weakens easterly (westerly) winds across the eastern Pacific in the lower (upper) atmosphere, which is clearly seen in the wind anomalies as depicted in Figure 1.3 (Figure 1.4). These anomalous circulation features clearly reflect anomalous Walker circulation related to the canonical El Niño episode (Webster et al., 1998).

**1.6 Monsoon convection propagation characteristics**

The summer monsoon 2015 is associated with two prominent northward propagating bands of anomalous convection from the equatorial Indian Ocean to the Indian subcontinent, one during the month of June and another during the end of July to the beginning of August (Figure 1.8a,d). The time-latitude section of daily OLR averaged over the Indian subcontinent clearly depicts the anomalous south to north propagation of convection, extending mainly from 5o S to 25o N (Figure 1.8a,d). Furthermore, the northward movement of convection is anomalously suppressed during the month of July and from the second week of August through September (Figure 1.8d), consistent with the anomalous reduction in rainfall during these months (Figure 1.1f,i,l). It is to be noted that the northward propagating convective cloud bands were less over the Arabian Sea compared to the BoB (Figure 1.8b,c). Intriguingly, the Arabian Sea witnessed strong northward propagation of anomalous convection during the month of June, and weakened propagation by the end of July (Figure 1.8b,e). On the other hand, strong northward propagation is observed over the BoB during the first half of June and August, with a weak propagation during the first half of July (Figure 1.8c,f). The comparison between the northward propagation of convection over the Indian subcontinent (Figure 1.8a,d), the Arabian Sea (Figure 1.8b,e) and the BoB (Figure 1.8c,f) suggest that the convection over the Indian subcontinent has a significant contribution from the BoB branch throughout the season, whereas the Arabian Sea contribution was mainly during the month of June.

The propagation of convection to the core monsoon region from the east is mainly associated with the westward propagation of synoptic scale convective systems that originate over the warm waters of the BoB and also from the remnants of the west Pacific typhoons (Sikka and Gadgil, 1980; Rajeevan, 1993). These propagation characteristics have been examined by averaging daily OLR over the latitudinal belt of 15-25o N (Figure 1.9). Clear westward propagation can be observed during the month of June, second half of July, first half of August and also during the end of September (Figure 1.9a,b), supporting the prominent contribution of synoptic scale disturbances to the rainfall during these periods (as described in Section 3.1 see Figure 3.1). The modelling and simulation aspects of meridional and zonal propagation of convection on intra-seasonal scale are mostly dealt in Section 5.2 of Chapter 5. Consistently, rainfall averaged over the core monsoon zone (Rajeevan et al., 2010) is also anomalously high during the month of June, the second half of July and the last two weeks of September (Figure 1.2). Spectral analysis of rainfall over the core monsoon region clearly depicts that the oscillations with approximately 8 day and 16 day periodicities are significantly dominant during the monsoon season (Figure 1.10a). The maximum variance in the oscillation with 8 days periodicity occurs during the second half of June to the first half of August (Figure 1.10b), when a considerable amount of rainfall occurs over the region (Figure 1.2). This therefore suggests that the significant contribution of rainfall is from the seven intense low pressure systems formed during the three month period (Figure 3.1). On the other hand, the anomalous rainfall during the second half of September (Figure 1.2) appears to be mainly associated with the 10-20 day periodicity (Figure 1.10b). This is supported by the occurrence of a depression over the BoB during the third week of September (Figure 3.1), manifested as the westward propagation of anomalous convection (Figure 1.9) and by the absence of northward propagation (Figure 1.8a,d). A slower 30 day mode was also present during early June to mid-July and towards the end of September, but was statistically insignificant. It is known that the westward propagating synoptic scale systems are the manifestation of oscillations with 10-20 day (faster mode) periodicity, whereas the northward propagation of convection from the equatorial Indian Ocean is the manifestation of 30–60 day (slower mode) oscillations (Sikka and Gadgil, 1980). Thus it is intriguing to note that the rainfall activity during the deficient monsoon season of 2015 is dominated by the faster mode of variability. However, this is in contrast to the previous deficient years, which are mainly dominated by the slower 30-60 day mode of variability, whereas the faster 10–20 day mode is generally dominant during normal years of Indian summer monsoon rainfall (Kripalani et al., 2004).

Table 1.1 Area weighted rainfall (mm) for the country as a whole during the summer monsoon season of 2015. Courtesy: IMD

| Months | Actual (mm) | Normal (mm) | % of long period mean |
|---|---|---|---|
| June | 189.5 | 163.6 | 115.8 |
| July | 241.9 | 289.2 | 83.7 |
| August | 204.2 | 261.3 | 78.1 |
| September | 131.4 | 173.4 | 76.0 |
| June to September | 767.0 | 887.5 | 86.4 |

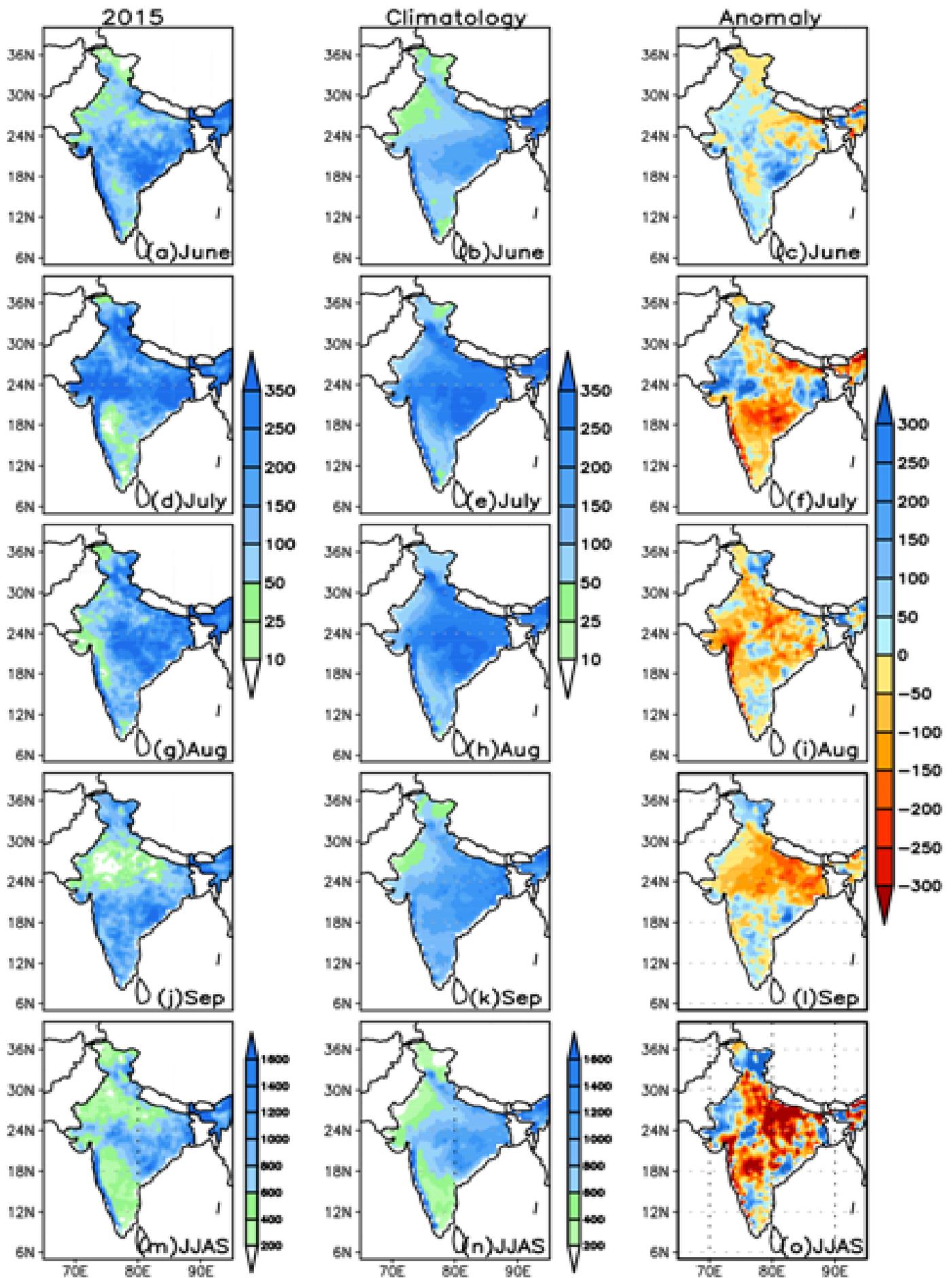

**Figure 1.1** Cumulative rainfall (mm) maps for the 2015 southwest monsoon. (a) Actual, (b) Climatology and (c) Anomalous rainfall for the month of June. (d)-(f) are the same, but for July. Similarly, (g)-(i) for August, (j)-(l) for September and (m)-(o) for June to September (JJAS) season. Gridded data are from www.imd.gov.in. The base period for climatology is 1980-2010, for all figures of this chapter.

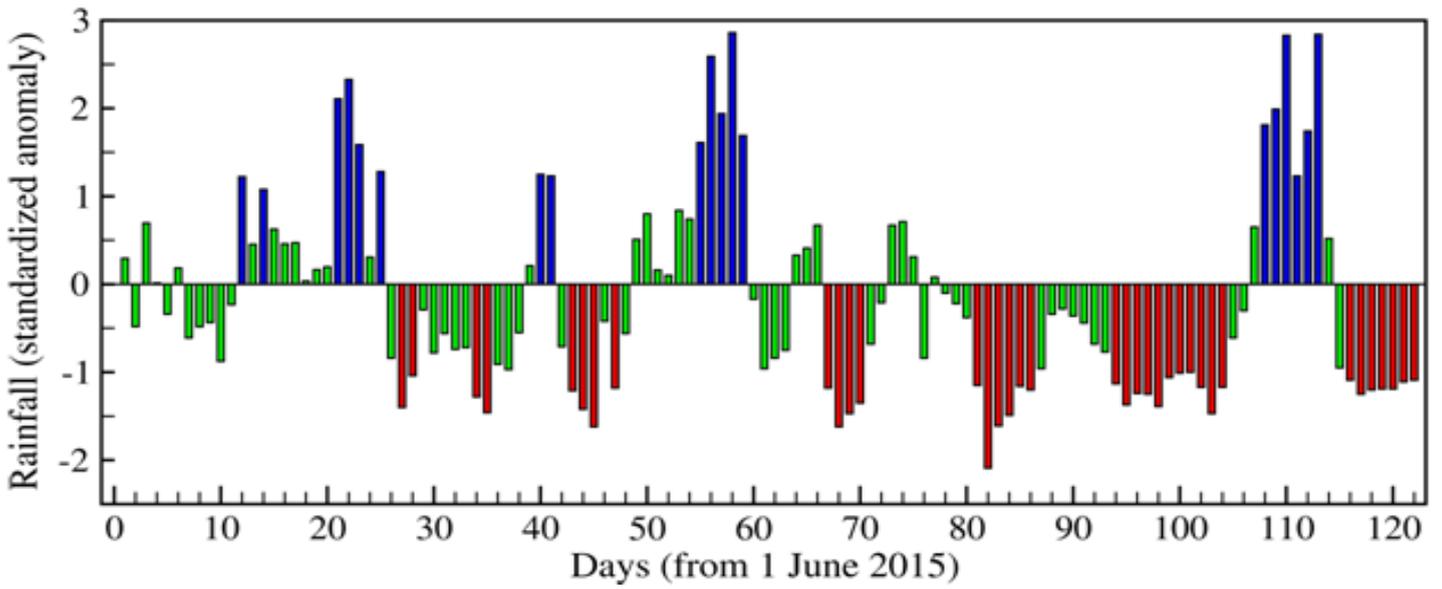

**Figure 1.2** Standardized daily rainfall anomaly over the core monsoon zone, for the summer monsoon 2015. Days with rainfall anomalies in red color indicates break monsoon days (Standardized anomaly less than -1.0) and blue colour indicates active monsoon days (Standardized anomaly more than +1.0).

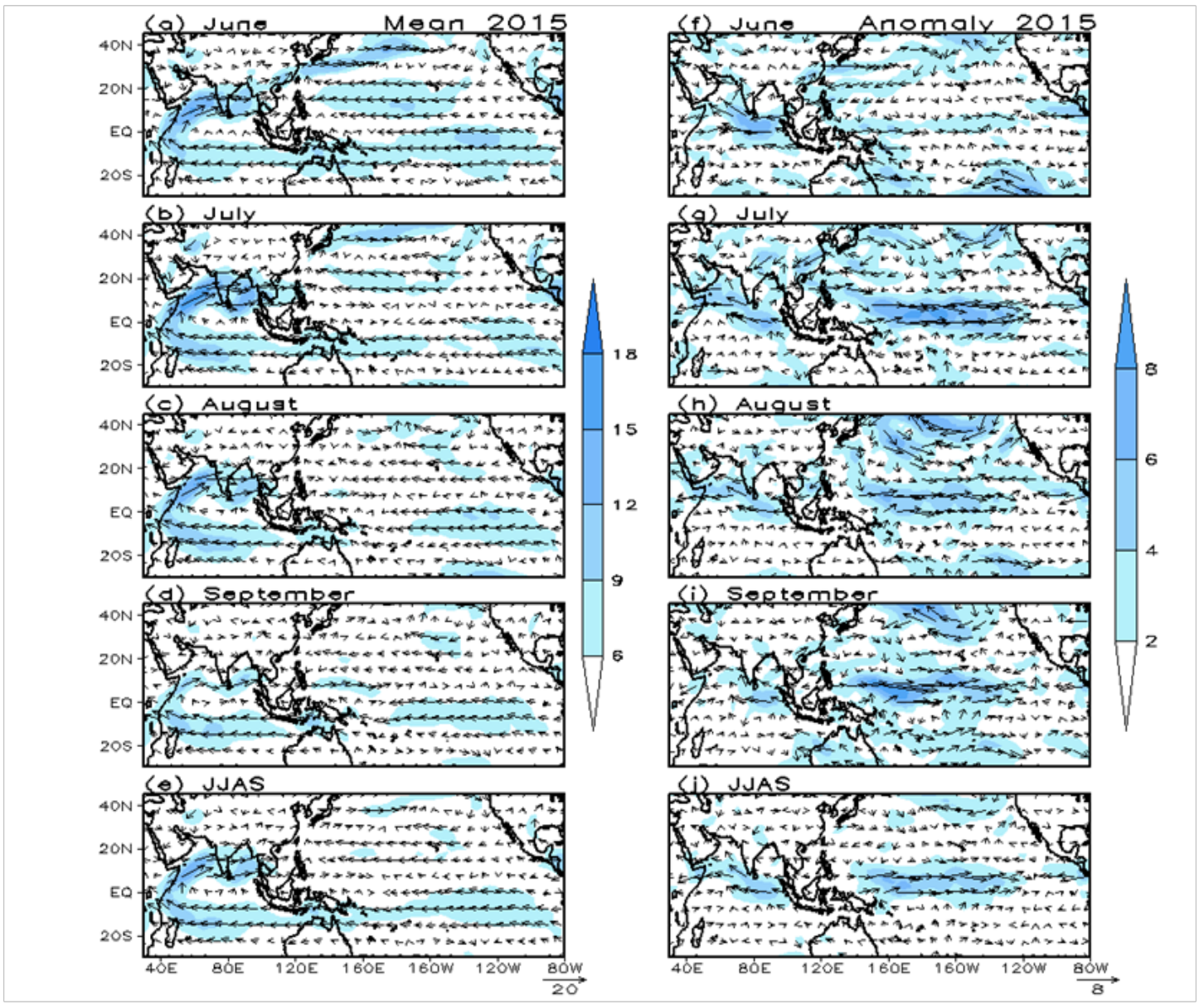

**Figure 1.3** (a-e) mean and (f-j) anomalies of winds (ms-1) at 850 hPa for (a,f) June, (b,g) July, (c,h) August, (d,i) September and (e,j) JJAS season during 2015. Data source: NCEP/NCAR reanalysis.

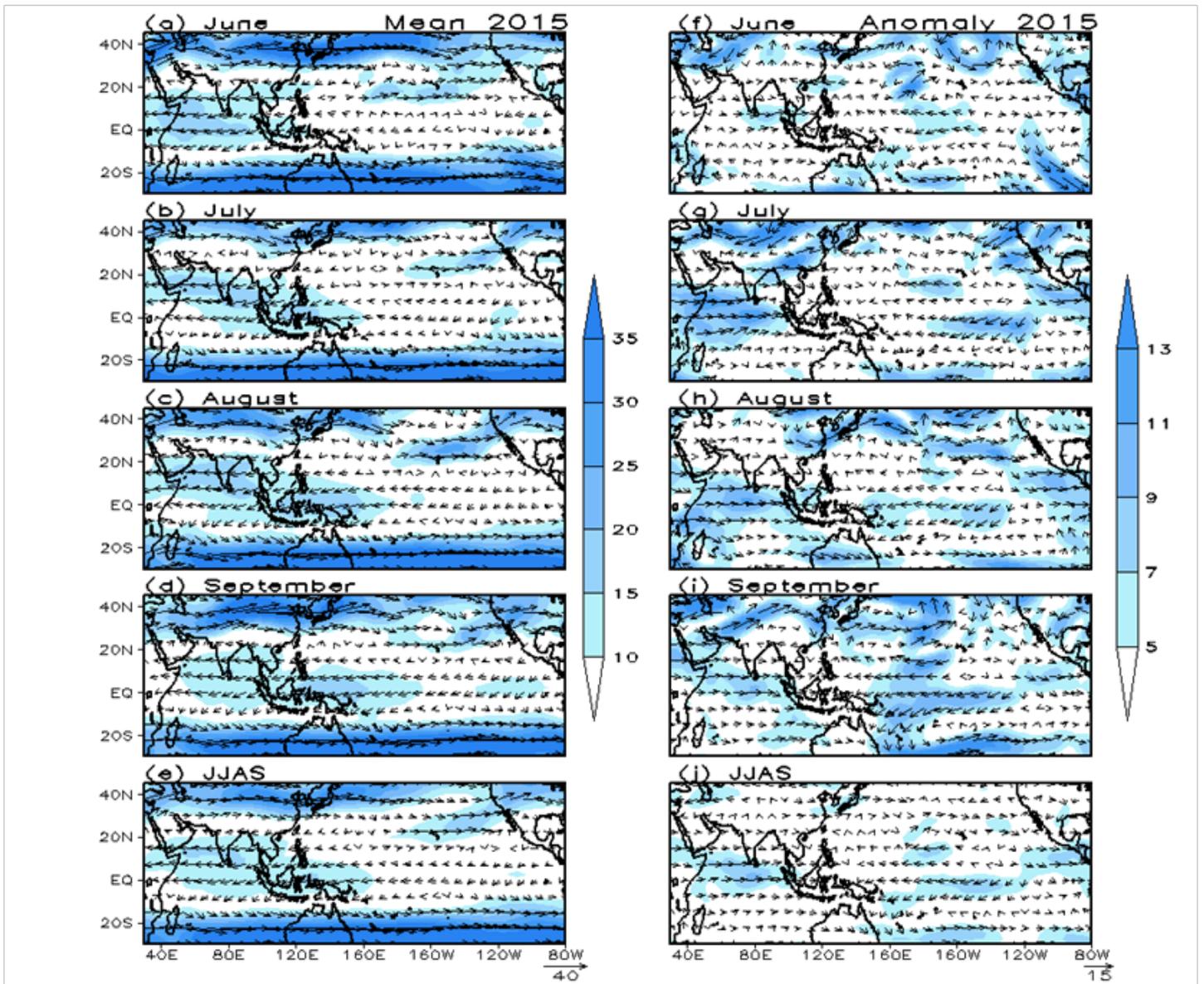
**Figure 1.4** Similar to Figure 1.2, except for 200 hPa.

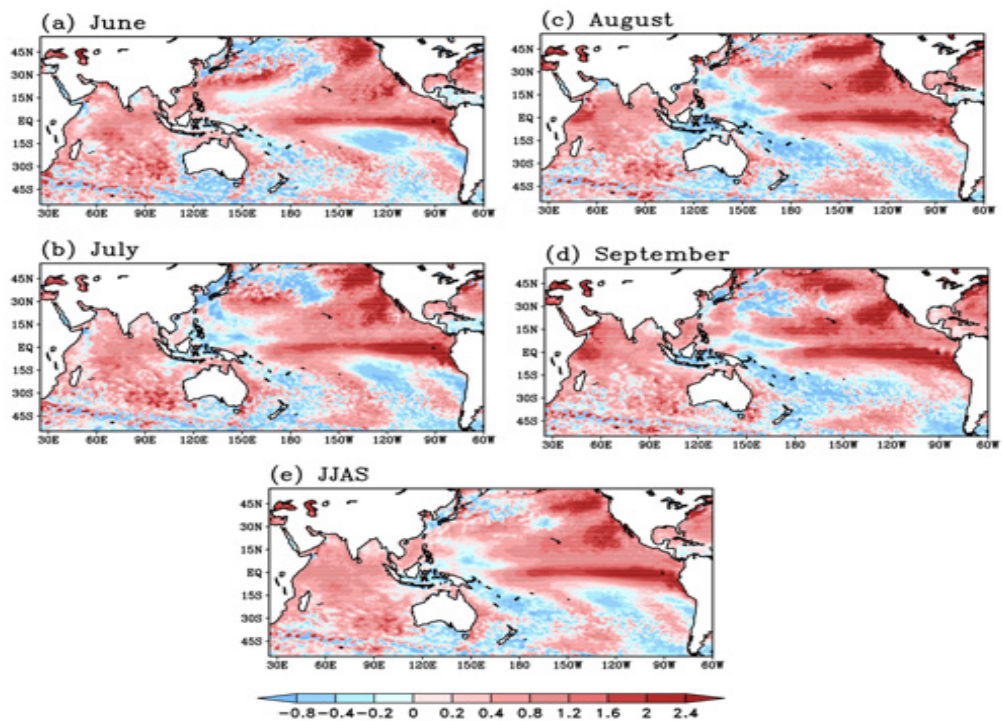
**Figure 1.5** Sea surface temperature anomalies (°C) for (a) June, (b) July, (c) August, (d) September and (e) JJAS season during 2015. Data source: Reynolds SST.

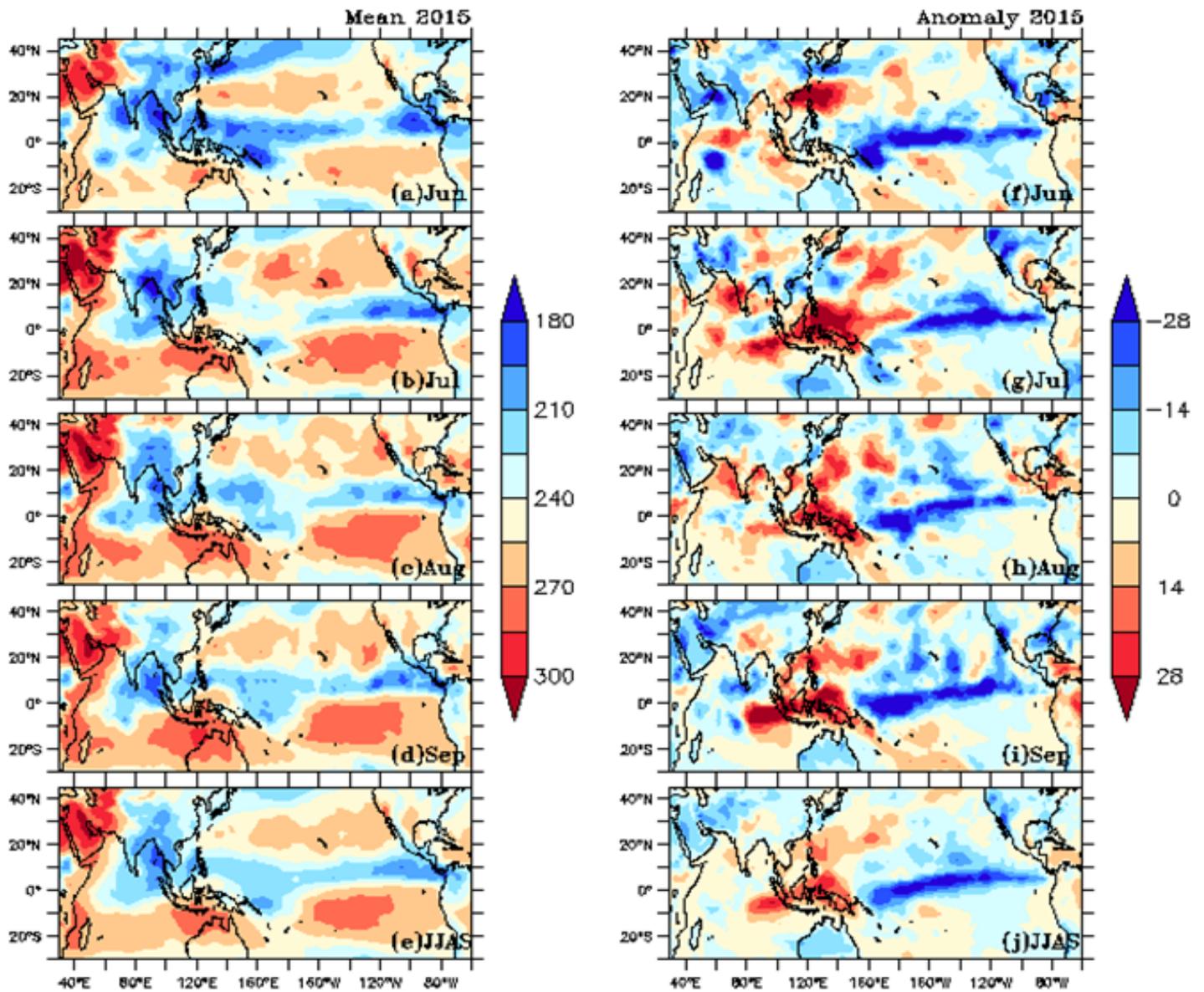

**Figure 1.6** Monthly mean (left panel) and anomalies (right panel) of OLR (Wm$^{-2}$) during 2015. Deep convection in the tropics is characterized by low cloud-top temperatures and small OLR values; while high OLR values indicate scarcity or absence of cloud cover. The spatial distributions of monthly mean OLR during the months of June to September are shown in (a) to (d), respectively. (e) Illustrates the JJAS seasonal mean of OLR. Right panel is similar to that of left panel, except for OLR anomalies. Low (high) / negative (positive) OLR values are blue (red) shaded. Data source: NOAA (http://www.cdc.noaa.gov).

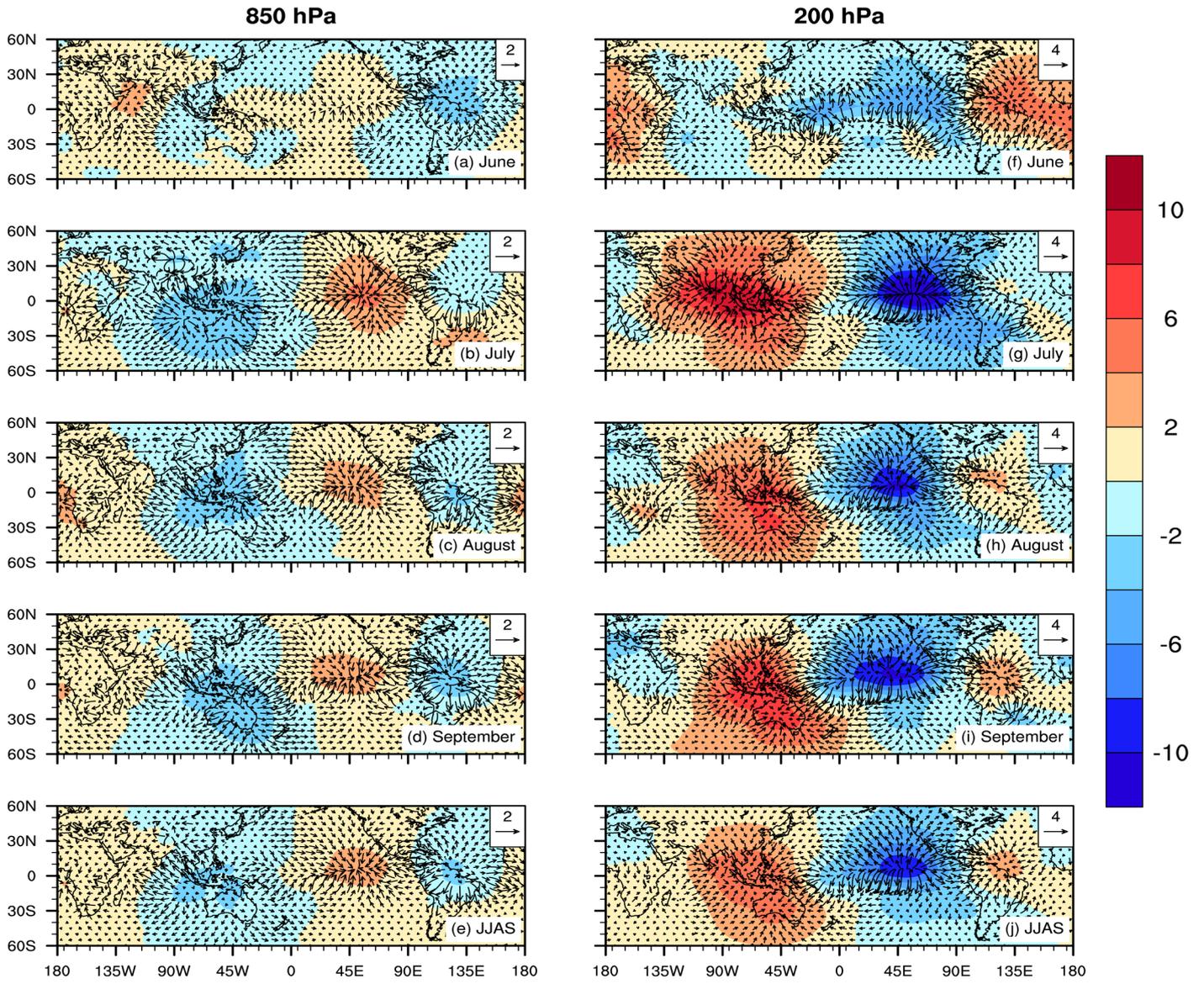

**Figure 1.7** Velocity Potential anomalies (106 m2/s) at (a-e) 850 hPa and (f-j) 200 hPa for the months of June, July, August, September and for the JJAS season. Data source: NCEP/NCAR reanalysis

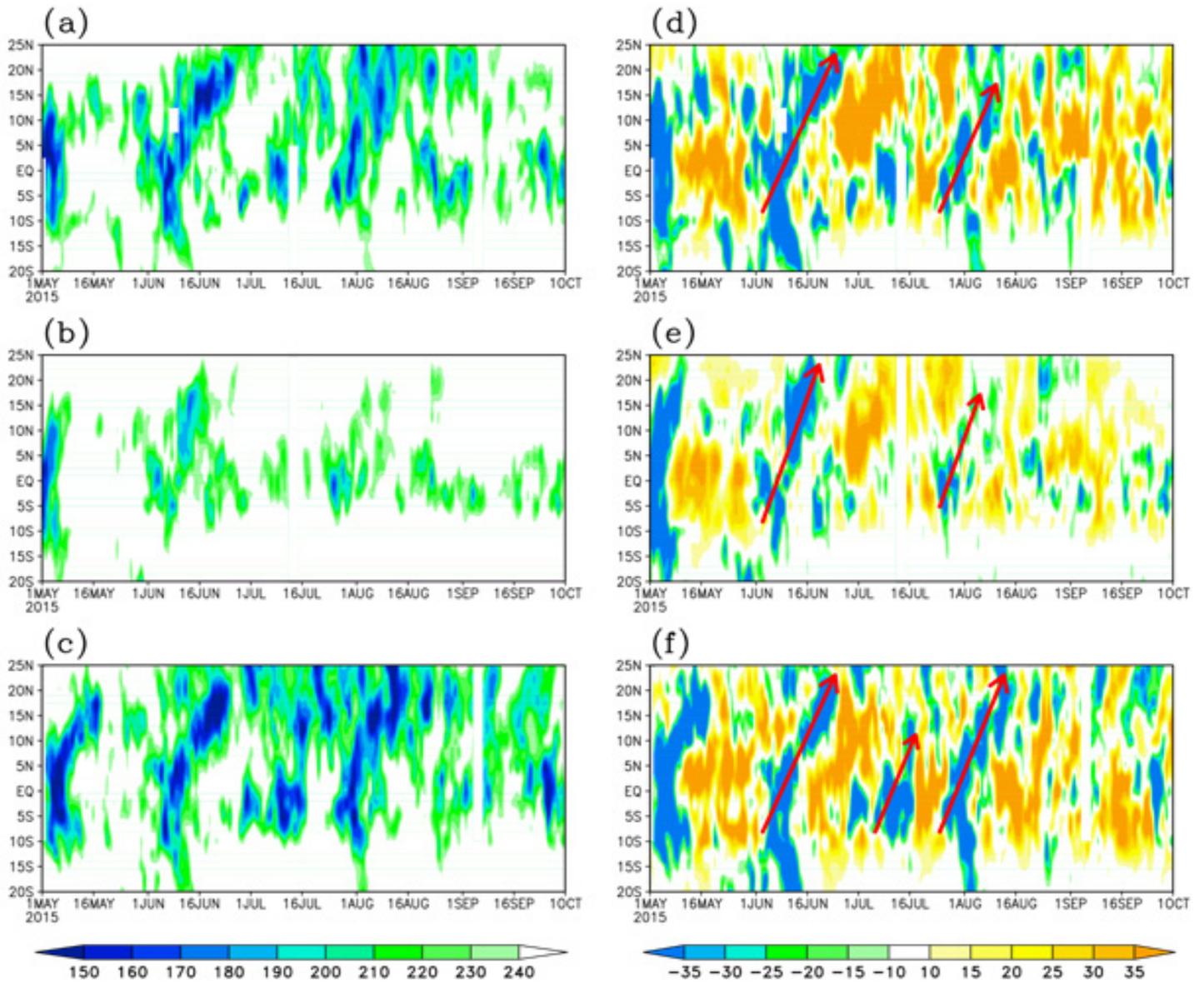

**Figure 1.8** Time-Latitude sections of (a-c) mean and (d-f) anomalies in daily OLR (Wm$^{-2}$) for the summer monsoon 2015. Panels (a and d) for the Indian subcontinent (averaged over the longitude belt of 70-85°E), (b and e) for the Arabian Sea (averaged over the longitude belt of 50-75°E) and (c and f) for the Bay of Bengal (averaged over the longitude belt of 80-95°E). Red arrows denote dominant northward propagating bands of anomalous convection. Data source: NOAA.

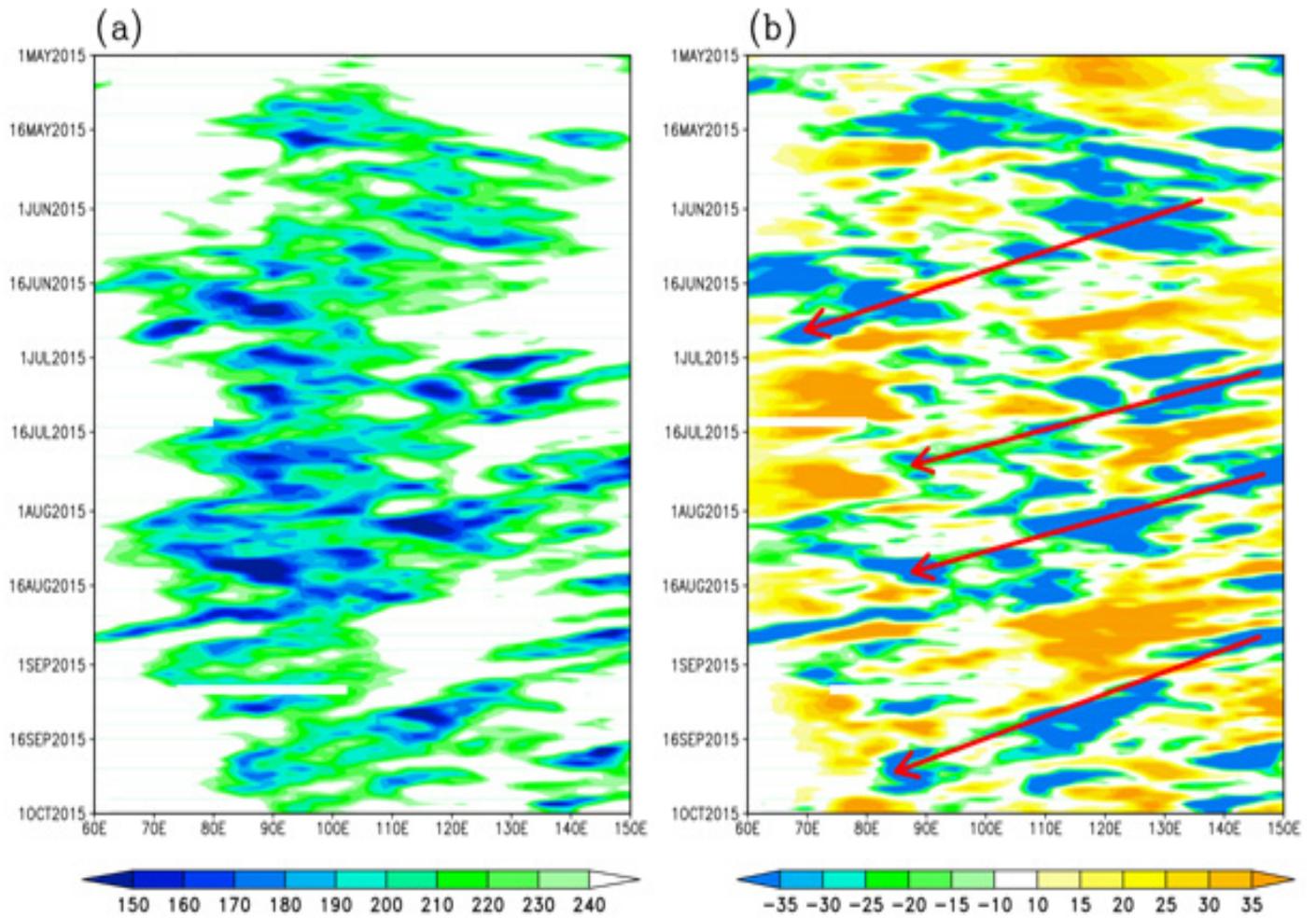

**Figure 1.9** Time-Longitude sections of (a) mean and (b) anomalies in daily OLR (Wm$^{-2}$) for the summer monsoon 2015, averaged over the latitude belt of 15-25°E. Red arrows denote dominant westward propagating bands of anomalous convection. Data source: NOAA

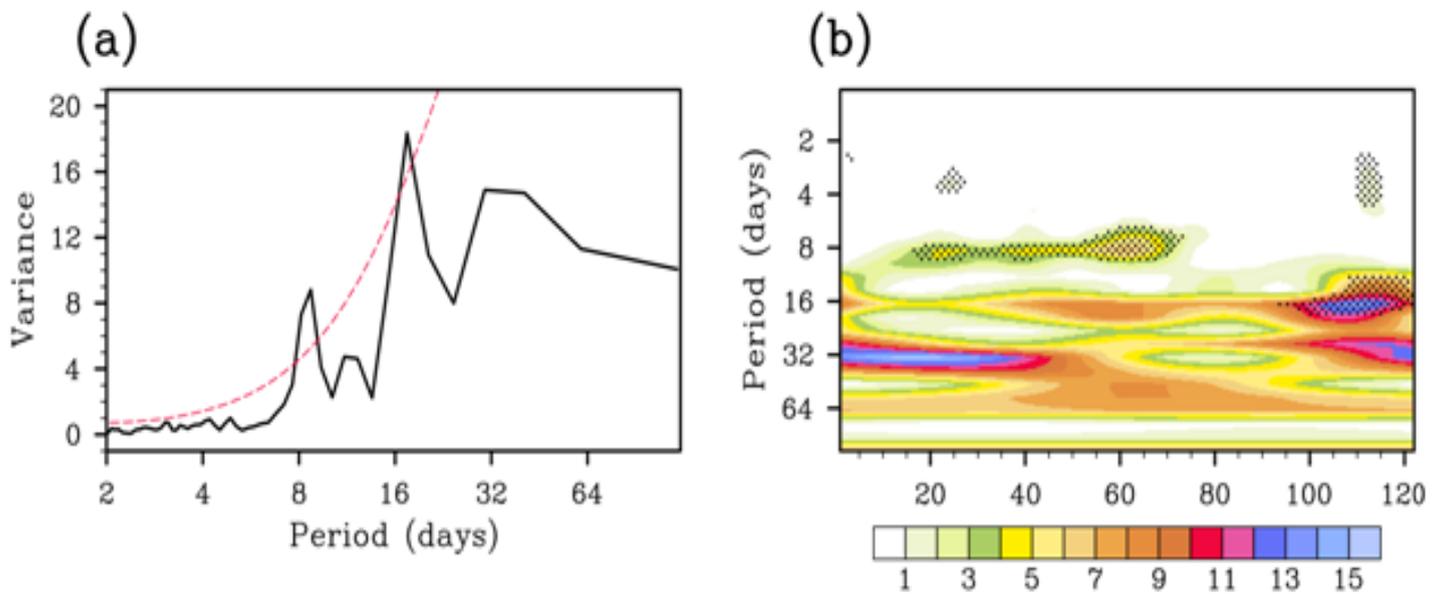

**Figure 1.10** (a) Spectral and (b) wavelet analysis of rainfall over the core monsoon zone, for the summer monsoon 2015. Red dash line in (b) and black dots in (c) denote statistical significance at 95% confidence level. Data source: IMD.